\documentclass[conference]{IEEEtran}

\IEEEoverridecommandlockouts
\def\BibTeX{{\rm B\kern-.05em{\sc i\kern-.025em b}\kern-.08em
		T\kern-.1667em\lower.7ex\hbox{E}\kern-.125emX}}

\usepackage{dblfloatfix}    
\usepackage{cite}
\usepackage{amsmath,amssymb,amsfonts}
\usepackage{bbm}

\usepackage{graphicx}
\usepackage{textcomp}
\usepackage{xcolor}
\usepackage{colortbl} 

\usepackage{url} 

\usepackage{bm}
\usepackage[normalem]{ulem} 
\usepackage{supertabular}
\usepackage{amsthm}
\theoremstyle{definition}

\usepackage[linesnumbered,ruled,vlined]{algorithm2e}
\usepackage{algorithmicx} 
\usepackage{pbox}
\usepackage{float} 
\usepackage{subcaption} 
\usepackage{epstopdf}

\usepackage{tikz}
\usetikzlibrary{positioning,arrows.meta,intersections}
\usetikzlibrary{calc}

\usepackage{boldline,multirow}
\usepackage{array}

\usepackage{pbox}

\usepackage{dblfloatfix}

\begin{document}
\setlength{\abovedisplayskip}{2pt}
\setlength{\belowdisplayskip}{2pt}


\title{Scalable Distortion-Aware Clustering for Fronthaul-Limited Cell-Free MIMO Networks 
}

\author{
\IEEEauthorblockN{
Zehua~Li$^{*}$, 
Raviraj~Adve$^{*}$, 
Israfil~Bahceci$^{\dagger}$, 
Yahia~Ahmed$^{\dagger}$
}
\IEEEauthorblockA{
$^{*}$\textit{Dept. of Electrical and Computer Engineering, University of Toronto}, Toronto, Canada 
}
\IEEEauthorblockA{
$^{\dagger}$\textit{Ericsson Canada}, Ottawa, Canada \\
Emails: samzehuali.li@mail.utoronto.ca, rsadve@ece.utoronto.ca, \{israfil.bahceci, yahia.ahmed\}@ericsson.com
}
}


%
%

\maketitle 


\begin{abstract}
This paper studies distortion-aware clustering for uplink fronthaul-limited cell-free MIMO networks employing maximum ratio combining (MRC). While MRC is appealing for its low-complexity, its performance is limited by interference, fronthaul distortions, and channel imperfections. This motivates clustering strategies that compensate for these impairments through coordination of access points. Since instantaneous small-scale fading-based optimization is not scalable in large systems, and is impractical due to frequent channel variations, we instead optimize an objective depending only on large-scale fading coefficients. To this end, asymptotic analysis is used to derive deterministic equivalent expressions for the average network sum rate, with focus on quantization distortion, leading to a quadratic-over-linear objective. Although maximizing such an objective under binary constraints is non-convex, we exploit its structure to develop a polynomial time scheme that attains the global optimum. Numerical results show that the proposed clustering method provides performance gains over improved variants of existing literature, and remains competitive with the small-scale fading-based global optimum obtained via exhaustive search.

\end{abstract}

%

%
%
%
\section{Introduction} \label{section:intro}
Cell-free multi-input multi-output (MIMO) networks are envisioned as a key technology for next-generation wireless systems~\cite{Björnson_2020}. By coordinating distributed access points (APs) through a central processing unit (CPU), cell-free architectures provide spatial diversity and improved channel conditions by bringing APs closer to users. Centralized processing can also improve interference management, for example, through minimum mean square error (MMSE) combining at the CPU. However, MMSE receivers do not scale well as the network densifies~\cite{Demir_2021,li_2024}, motivating clustering methods where each user is served by only a subset of APs.

A low complexity alternative is maximum ratio combining (MRC), which is more scalable but lacks an interference suppression capability. This makes cluster selection crucial, since better clusters can reduce interference and improve user rates~\cite{jayawardane_2025}. The problem becomes even more challenging in fronthaul-limited networks, where APs must compress their received signals before forwarding them to the CPU, introducing quantization distortion~\cite{Yu_Patil_Dai_Zhou_2017,Liu_Yu_Simeone_2017}. Since MRC does not explicitly account for interference or fronthaul distortion, clustering for MRC receivers requires a design perspective different from clustering for MMSE receivers.

Prior works have separately studied clustering design~\cite{jayawardane_2025} and fronthaul limitations~\cite{li_2026}, often optimizing objectives that depend on small-scale fading information. While such objectives provide accurate performance characterization, they require frequent updates. Practical aspects have also been explored, including learning-based clustering~\cite{al_2021} and partial use of channel statistics for fronthaul compression~\cite{li_2025}. However, these methods still rely on small-scale fading information and are therefore less scalable under rapidly varying channels.

A more scalable approach is to rely only on large-scale fading coefficients (LSFCs), so that optimization is required only at the time scale of LSFC variations. One established method is to use asymptotic analysis to derive deterministic equivalent (DE) expressions for average rates and optimize these approximations instead~\cite{wagner_2012}. While DE-based optimization incurs some performance loss by targeting average rate approximations rather than instantaneous rates, it is more practical because the resulting strategies require far less frequent updates. We emphasize that only clustering is updated at the LSFC time
scale, while MRC weights are still formed using small-scale fading realizations.

The main focus of this paper is scalable distortion-aware clustering for fronthaul-limited cell-free MIMO networks with MRC receivers. The main contributions are:
\begin{itemize}
    \item We formulate the instantaneous achievable user rate under limited fronthaul and imperfect channel state information.
    \item We derive a DE-based rate expression that accounts for MRC and quantization distortion. This leads to a per-user cluster selection problem in the form of a quadratic-over-linear maximization under binary constraints. Despite this being non-convex, we derive a globally optimal solution.
    \item Through simulations, we validate the DE approximation, verify the LSFC-based global optimality of the proposed clustering scheme, and compare against exhaustive search and enhanced versions of literature benchmarks~\cite{jayawardane_2025} that account for limited fronthaul.
\end{itemize}

\section{System Model}\label{section:system_model}
\subsection{Channel and Signal Model}\label{subsection:channel_and_signal_model}
We consider the uplink of a cell-free MIMO network operating in time-division duplex (TDD) mode. A single CPU controls a set of APs denoted by the set $\mathcal{R}$; each AP is equipped with $M$ antennas. The set of single-antenna users (UEs) is denoted as $\mathcal{U}$. We use subscripts $r$ and $u$ to refer to APs and users respectively. We refer to the links between users and APs as access links and between APs and CPU as fronthaul links. Each fronthaul link has a spectral efficiency (SE) limit of ${L}_r$ bits/s/Hz. We refer to rate and SE interchangeably since, in our TDD mode, they only differ by a scale factor.

For the access channel $\mathbf{h}_{ru}$ between user $u$ and AP $r$, only an imperfect channel estimate $\hat{\mathbf{h}}_{ru}$ is available. From~\cite{ammar_2022},
\begin{equation}\label{eqn:channel_model}
\begin{aligned}
    \mathbf{h}_{ru} &= \hat{\mathbf{h}}_{ru} + \mathbf{e}_{ru} 
    = \mathbf{\Psi}_{ru}^{1/2} \mathbf{g}_{ru} + 
    \tilde{\mathbf{\Psi}}_{ru}^{1/2} \tilde{\mathbf{g}}_{ru} ,
\end{aligned}
\end{equation}
where $\mathbf{e}_{ru}$ denotes estimation error, $\mathbf{\Psi}_{ru}$ denotes the channel correlation matrix, and $\tilde{\mathbf{\Psi}}_{ru}$ denotes the error covariance. $\mathbf{g}_{ru}, \tilde{\mathbf{g}}_{ru} \sim \mathcal{CN}(\mathbf{0}_{M\times 1},\mathbf{I}_M)$ are the small-scale fading and normalized estimation error assumed mutually independent.

The user transmission assumes Gaussian signaling and is denoted as $x_u\sim\mathcal{CN}(0,p_u)$ where $p_u$ is the transmit power. At AP $r$, the received signal vector $\mathbf{y}_r\in\mathbb{C}^M$ is given by
\begin{equation}
    \mathbf{y}_r = \sum_{u\in\mathcal{U}}\mathbf{h}_{ru} x_u + \mathbf{n}_r 
    = \sum_{u\in\mathcal{U}} ( \hat{\mathbf{h}}_{ru}+\mathbf{e}_{ru} ) x_u + \mathbf{n}_r ,
\end{equation}
where $\mathbf{n}_r \sim \mathcal{CN}(\mathbf{0}_{M\times 1},\mathbf{\Sigma}_{\mathbf{n}_r})$ denotes the additive Gaussian noise. The APs then compress (quantize) the received signals and forward the compressed versions to CPU. This is referred as compress-forward~\cite{Yu_Patil_Dai_Zhou_2017} or centralized operation~\cite{Demir_2021}. The compressed signal vector $\boldsymbol{\zeta}_r\in\mathbb{C}^M$ seen at CPU is given by
\begin{equation}
    \boldsymbol{\zeta}_r = \mathcal{Q} ( \mathbf{y}_r ) = \mathbf{y}_r + \mathbf{q}_r
    = \sum_{u\in\mathcal{U}} ( \hat{\mathbf{h}}_{ru}+\mathbf{e}_{ru} ) x_u + \mathbf{n}_r + \mathbf{q}_r ,
\end{equation}
where $\mathcal{Q}(\cdot)$ denotes the quantization operation. We assume a Gaussian test channel model~\cite{Yu_Patil_Dai_Zhou_2017,Liu_Yu_Simeone_2017,liu_2015} with quantization error $\mathbf{q}_r \sim \mathcal{CN}(\mathbf{0}_{M\times 1},\mathbf{\Sigma}_{\mathbf{q}_r})$. The covariance $\mathbf{\Sigma}_{\mathbf{q}_r}$ is designed to satisfy the fronthaul limit and will be discussed later.

At the CPU, only signals from a subset of APs, called user's serving cluster, are combined. For each user $u$, we define $\mathcal{C}_u \subseteq \mathcal{R}$ as its serving cluster. 
The estimated symbol yields
\begin{align} \label{eqn:estimate_symbol}
	\hat{x}_u &= \sum_{r \in \mathcal{C}_u} \mathbf{w}_{ru}^{\mathsf{H}} \boldsymbol{\zeta}_r  \notag \\
	&= \mathbf{w}_{\mathcal{C}_{u}}^{\mathsf{H}} \left( \sum_{u' \in \mathcal{U}} ( \hat{\mathbf{h}}_{\mathcal{C}_{u}u'} + \mathbf{e}_{\mathcal{C}_{u}u'} ) x_{u'} + \mathbf{n}_{\mathcal{C}_{u}} + \mathbf{q}_{\mathcal{C}_{u}} \right) ,
\end{align}
where $\mathbf{w}_{ru} \in \mathbb{C}^{M}$ is the combiner for user $u$ corresponding to the signals from AP $r\in\mathcal{C}_u$. We introduce using $\mathcal{C}_{u}$ as subscripts to denote vertical concatenations of all vectors from the set as $\mathbf{w}_{\mathcal{C}_{u}}, \mathbf{n}_{\mathcal{C}_{u}}, \mathbf{q}_{\mathcal{C}_{u}}, \hat{\mathbf{h}}_{\mathcal{C}_{u}u'}, \mathbf{e}_{\mathcal{C}_{u}u'} \in \mathbb{C}^{M|\mathcal{C}_u|}$. For example, $\hat{\mathbf{h}}_{\mathcal{C}_{u}u'}$ is the collective estimated channels between cluster $\mathcal{C}_u$ and user $u'$, i.e., $\hat{\mathbf{h}}_{\mathcal{C}_{u}u'} = [ \cdots,
\hat{\mathbf{h}}_{ru'}^{\mathsf{T}}, \cdots  ]_{r\in\mathcal{C}_u}^{\mathsf{T}}$. $\mathbf{w}_{\mathcal{C}_{u}}$ is viewed as the centralized receive combiner used at the CPU to estimate the symbol sent by user $u$. The collective noise and quantization distortion covariances are $\mathbf{\Sigma}_{\mathbf{n}_{\mathcal{C}_{u}}}=\mathrm{diag}\{\mathbf{\Sigma}_{\mathbf{n}_r}\}_{r\in\mathcal{C}_u}$ and $\mathbf{\Sigma}_{\mathbf{q}_{\mathcal{C}_{u}}}=\mathrm{diag}\{\mathbf{\Sigma}_{\mathbf{q}_r}\}_{r\in\mathcal{C}_u}$. Similarly, the collective channel correlation matrix is $\mathbf{\Psi}_{\mathcal{C}_{u}u'}=\mathrm{diag}\{\mathbf{\Psi}_{ru'}\}_{r\in\mathcal{C}_u}$ and similarly for the collective error covariance $\tilde{\mathbf{\Psi}}_{\mathcal{C}_{u}u'}$.

The instantaneous effective signal-to-interference-plus-quantization-and-noise ratio (SIQNR), similar to~\cite{Demir_2021,li_2026}, is
{
\allowdisplaybreaks
\begin{align}
    \hspace{-0.41em}
    \mathsf{SIQNR}_u &= \frac{p_u | \mathbf{w}_{\mathcal{C}_{u}}^{\mathsf{H}} \hat{\mathbf{h}}_{\mathcal{C}_{u}u}  |^2}
	{ \mathsf{EN}_u + \mathsf{Intf}_u
    }  = \gamma_u  ,  \label{eqn:siqnr}
    \\
    \mathsf{EN}_u &=  \mathbf{w}_{\mathcal{C}_{u}}^{\mathsf{H}} ( \mathbf{\Sigma}_{\mathbf{n}_{\mathcal{C}_{u}}} + \mathbf{\Sigma}_{\mathbf{q}_{\mathcal{C}_{u}}} +
    \textstyle\sum_{u'\in\mathcal{U}}  p_{u'} \tilde{\mathbf{\Psi}}_{\mathcal{C}_{u}u'} 
    ) \mathbf{w}_{\mathcal{C}_{u}}  
    , \\
    \mathsf{Intf}_u &= \textstyle\sum_{u'\in\mathcal{U}\backslash u} p_{u'} | \mathbf{w}_{\mathcal{C}_{u}}^{\mathsf{H}} \hat{\mathbf{h}}_{\mathcal{C}_{u}u'} |^2  ,
\end{align}
}
where $\mathsf{Intf}_u$ denotes the contribution of interference and $\mathsf{EN}_u$ is the effective noise that includes quantization distortions and channel uncertainties. Then, an achievable rate in centralized operation is $R_u=\log_2(1+\mathsf{SIQNR}_u)=\log_2(1+\gamma_u)$.

\subsection{Compression Model}\label{subsection:compression_model}
The codebook design of the quantizers is equivalent to setting the covariances of quantization noise $\mathbf{\Sigma}_{\mathbf{q}_r}$~\cite{Liu_Yu_Simeone_2017,Yu_Patil_Dai_Zhou_2017}. The distortion should be made as small as possible to achieve higher rates where the minimum amount of distortion is limited by fronthaul capacity, given by rate-distortion theory. 

Similar to~\cite{Liu_Yu_Simeone_2017,li_2026}, the rate (normalized over access bandwidth) required to transmit the signal received at the $m$-th antenna of AP $r$ over the fronthal channel is given by
\begin{align}
    l_{r,m} &= \log_2 \left(
    \frac{ \Gamma_q \sigma_{y_{r,m}}^2 +  \sigma^2_{q_{r,m}} }
    {\sigma^2_{q_{r,m}}}
    \right)  , \\
    \sigma_{y_{r,m}}^2 &= \left[ \textstyle\sum\nolimits_{u\in\mathcal{U}} p_{u} ( \hat{\mathbf{h}}_{ru}\hat{\mathbf{h}}_{ru}^{\mathsf{H}} + \tilde{\mathbf{\Psi}}_{ru} ) + \mathbf{\Sigma}_{\mathbf{n}_r} \right]_{mm} ,
\end{align}
where $[\cdot]_{ij}$ is to denote $(i,j)$-th entry of a matrix and $\Gamma_q$ is the gap to rate-distortion theory. For simplicity, the total fronthaul rate is allocated equally to its $M$ antennas, i.e. $l_{r,m}=L_r/M,\forall r$. The distortion variance is then
\begin{equation}\label{eqn:rate_distortion_relation}
    \sigma^2_{q_{r,m}} = \frac{\Gamma_q \sigma_{y_{r,m}}^2}{2^{L_r/M}-1} ,
\end{equation}
for each antenna and the distortion covariance for each AP is $\mathbf{\Sigma}_{\mathbf{q}_r}=\mathrm{diag}\{\sigma^2_{q_{r,m}}\}_{m=1}^M$. By defining $\kappa_r = \Gamma_q/(2^{L_r/M}-1)$ and $\mathbf{K}_r^{\mathfrak{D}}=\kappa_r\mathbf{I}_M$, the distortion covariance can also be alternatively written as
\begin{equation}
    \mathbf{\Sigma}_{\mathbf{q}_r} = \mathbf{K}_r^{\mathfrak{D}} \odot 
    \left(
    \textstyle\sum\nolimits_{u\in\mathcal{U}} p_{u} ( \hat{\mathbf{h}}_{ru}\hat{\mathbf{h}}_{ru}^{\mathsf{H}} + \tilde{\mathbf{\Psi}}_{ru} ) + \mathbf{\Sigma}_{\mathbf{n}_r}
    \right)   ,
\end{equation}
with $\odot$ denoting the Hadamard product. The superscript-${\mathfrak{D}}$ is used on matrices to remind us that they are diagonal. The intention of $\mathbf{A}^{\mathfrak{D}}\odot\mathbf{B}$ is to only keep the diagonal of the entry-wise multiplication of the two matrices.

We choose to use the Gaussian test channel model as~\eqref{eqn:rate_distortion_relation} provides continuous relationships between rate of quantization and distortion. Other models such as Bussgang decomposition~\cite{bussgang_1952} or AQNM~\cite{max_1960} can also be extended to our analysis.

We remark that the APs may alternatively perform local combining before forwarding to the CPU~\cite{Demir_2021}; however, this makes the distortion in~\eqref{eqn:rate_distortion_relation} dependent varies with $|\mathcal{C}_u|$ or cluster-dependent. The centralized scheme avoids this coupling and yields the tractable clustering formulations in Section~\ref{section:optimal_de_based_clustering}, while MRC reduces complexity of centralized processing.

\section{Asymptotic Analysis of SIQNR under MRC}
For this paper, maximum ratio combining (MRC) is used for centralized processing where $\mathbf{w}_{\mathcal{C}_u} = \hat{\mathbf{h}}_{\mathcal{C}_u u}$. Compared to zero-forcing and MMSE combiner, the data rates achieved when using MRC degrade as it lacks interference management. However, especially in dense networks, it is less complex to implement making it more practical.

As a first step of our development, we use the large system approach and derive the asymptotic SIQNR under MRC. More precisely, we approximate the SIQNR of user $u$ by a deterministic equivalent (DE) $\gamma_u^o$ such that $\gamma_u-\gamma_u^o\rightarrow 0$ almost surely, as the system dimensions $M$ and $|\mathcal{U}|$ go to infinity. The approximation is tight asymptotically and yields accurate results even when the dimensions are small~\cite{wagner_2012}.


There are some common assumptions to be made for the asymptotic analysis. The system dimension ratio is bounded, i.e. $0<\lim_{M,|\mathcal{U}|\rightarrow \infty} M/|\mathcal{U}| < \infty$, and the correlation matrices $\mathbf{\Psi}_{ru},\tilde{\mathbf{\Psi}}_{ru}$ have uniformly bounded spectral norm. The per-antenna compression rates are assumed to be strictly positive, i.e. $\lim_{M\rightarrow\infty}L_r/M > 0$, to avoid the degenerate case of no service. To denote all system dimensions approach infinity with the above assumptions, we use $\mathcal{M}\rightarrow\infty$ and denote $a-a^o\overset{\mathcal{M}\rightarrow\infty}{\longrightarrow}0$ almost surely as $a\asymp a^o$.

For each user, let $M_u=M|\mathcal{C}_u|$ denote the combining dimension. We define the normalized small-scale fading vector $\mathbf{z}_{\mathcal{C}_u u'}\sim\mathcal{CN}(\mathbf{0}_{M_u\times1},\frac{1}{M_u}\mathbf{I}_{M_u})$ and normalized channels $\mathbf{b}_{\mathcal{C}_u u'}$ such that $\hat{\mathbf{h}}_{\mathcal{C}_u u'} = \sqrt{M_u} \mathbf{\Psi}_{\mathcal{C}_u u'}^{1/2} \mathbf{z}_{\mathcal{C}_u u'}=\sqrt{M_u}\mathbf{b}_{\mathcal{C}_u u'}$. 

To find the DE expressions for SIQNR under MRC, we introduce the following lemmas. 

\textit{Lemma 1:} For each positive integer $N$, let $\mathbf{A}_N\in\mathbb{C}^{N\times N}$ be a random matrix with uniformly bounded spectral norm. Let $\mathbf{x}_N\in\mathbb{C}^{N}$ be a random vector of i.i.d. entries with zero mean, variance $1/N$, and independent of $\mathbf{A}_N$. Then
\begin{equation}
    \mathbf{x}_N^{\mathsf{H}} \mathbf{A}_n \mathbf{x}_N 
    - \frac{1}{N} \mathrm{tr} \mathbf{A}_N
    \overset{N\rightarrow\infty}{\longrightarrow}  0  ,
\end{equation}
almost surely.

\textit{Proof:} The proof is given in~\cite{wagner_2012}. \hfill $\blacksquare$

\textit{Lemma 2:} Let $N$ and $\mathbf{x}_N$ be as in Lemma 1 and let $\mathbf{A}_N^{\mathfrak{D}},\mathbf{B}_N\in\mathbb{C}^{N\times N}$ be deterministic matrices with uniformly bounded spectral norm. Further, $\mathbf{A}_N^{\mathfrak{D}}$ is diagonal with positive entries and $\mathbf{B}_N$ is positive semidefinite. Then
\begin{equation}
\begin{split}
    \mathbf{x}_N^{\mathsf{H}} \mathbf{B}_N^{1/2} (\mathbf{A}_N^{\mathfrak{D}} \odot  (\mathbf{B}_N^{1/2} & \mathbf{x}_N \mathbf{x}_N^{\mathsf{H}}  \mathbf{B}_N^{1/2} ) ) \mathbf{B}_N^{1/2} \mathbf{x}_N - \\
    & \frac{2}{N^2} \mathrm{tr} \mathbf{B}_N ( \mathbf{A}_N^{\mathfrak{D}} \odot \mathbf{B}_N ) 
    \overset{N\rightarrow\infty}{\longrightarrow}  0  ,
\end{split}
\end{equation}
almost surely.

\textit{Proof:} The proof is given in Appendix~\ref{appendix:lemma_2}.   \hfill $\blacksquare$

The SIQNR and the corresponding DE is expressed as
{
\allowdisplaybreaks
\begin{align}
    \gamma_u &= \frac{\mathsf{Sgnl}_u}
    {\mathsf{Ns}_u+\mathsf{Impf}_u+\mathsf{Intf}_u+\sum_{i=1}^4\mathsf{QN}_{i,u}}  \notag   \\
    \label{eqn:de_siqnr} &\asymp \frac{\mathsf{Sgnl}_u^o} 
    {\mathsf{Ns}_u^o+\mathsf{Impf}_u^o+\mathsf{Intf}_u^o+\sum_{i=1}^4\mathsf{QN}_{i,u}^{o}} = \gamma_u^o    ,   \\   
    \mathsf{Sgnl}_u &= p_u | \mathbf{b}_{\mathcal{C}_{u}u}^{\mathsf{H}} \mathbf{b}_{\mathcal{C}_{u}u}  |^2  \notag \\
    &\asymp \frac{1}{M_u^2} ( \textstyle\sum\nolimits_{r\in\mathcal{C}_u} \mathrm{tr} \mathbf{\Psi}_{ru} )^2 =  \mathsf{Sgnl}_u^o ,   \\  
    \mathsf{Ns}_u &= \frac{1}{M_u} \mathbf{b}_{\mathcal{C}_{u}u}^{\mathsf{H}} \mathbf{\Sigma}_{\mathbf{n}_{\mathcal{C}_{u}}} \mathbf{b}_{\mathcal{C}_{u}u}  \notag  \\
    &\asymp \frac{1}{M_u^2} \textstyle\sum\nolimits_{r\in\mathcal{C}_u} \mathrm{tr} \mathbf{\Psi}_{ru} \mathbf{\Sigma}_{\mathbf{n}_{r}} =  \mathsf{Ns}_u^o ,  \\
    \mathsf{Impf}_u &= \frac{1}{M_u} \sum_{u'\in\mathcal{U}} p_{u'} \mathbf{b}_{\mathcal{C}_{u}u}^{\mathsf{H}} \tilde{\mathbf{\Psi}}_{\mathcal{C}_{u}u'} \mathbf{b}_{\mathcal{C}_{u}u}  \notag  \\
    &\asymp \sum_{u'\in\mathcal{U}} \frac{p_{u'}}{M_u^2} 
    ( \textstyle\sum_{r\in\mathcal{C}_u} \mathrm{tr} \mathbf{\Psi}_{ru} \tilde{\mathbf{\Psi}}_{ru'} ) = \mathsf{Impf}_u^o
    ,   \\
    \mathsf{Intf}_u &= \sum_{u' \in \mathcal{U} \backslash u} p_{u'} | \mathbf{b}_{\mathcal{C}_{u}u}^{\mathsf{H}} \mathbf{b}_{\mathcal{C}_{u}u'}  |^2 \notag   \\ 
    &\asymp \textstyle{ \sum_{u'\in\mathcal{U}\backslash u} 
    \frac{p_{u'}}{M_u^2}
    ( \textstyle\sum_{r\in\mathcal{C}_u} \mathrm{tr} \mathbf{\Psi}_{ru} \mathbf{\Psi}_{ru'} ) = \mathsf{Intf}_u^o  }
    ,   \\
    \mathsf{QN}_{1,u} &= \frac{1}{M_u}
    \mathbf{b}_{\mathcal{C}_u u}^{\mathsf{H}} 
    (
    \mathbf{K}_{\mathcal{C}_u}^{\mathfrak{D}} \odot \mathbf{\Sigma}_{\mathbf{n}_{\mathcal{C}_{u}}} 
    )
    \mathbf{b}_{\mathcal{C}_u u}   \notag  \\
    &\asymp \frac{1}{M_u^2} \textstyle\sum_{r\in\mathcal{C}_u} \mathrm{tr} 
    \mathbf{\Psi}_{ru} \left(
    \mathbf{K}_{r}^{\mathfrak{D}} \odot \mathbf{\Sigma}_{\mathbf{n}_{r}} 
    \right)  = \mathsf{QN}_{1,u}^o   ,    \\
    \mathsf{QN}_{2,u} &= \frac{1}{M_u} \sum_{u'\in\mathcal{U}} p_{u'} \mathbf{b}_{\mathcal{C}_{u}u}^{\mathsf{H}} 
    (
    \mathbf{K}_{\mathcal{C}_u}^{\mathfrak{D}} \odot \tilde{\mathbf{\Psi}}_{\mathcal{C}_{u}u'} 
    )
    \mathbf{b}_{\mathcal{C}_{u}u}  \notag  \\
    &\asymp \sum_{u'\in\mathcal{U}} \frac{p_{u'}}{M_u^2}  \sum_{r\in\mathcal{C}_u} 
    \mathrm{tr} \mathbf{\Psi}_{ru} (
    \mathbf{K}_{r}^{\mathfrak{D}} \odot \tilde{\mathbf{\Psi}}_{ru'} 
    )   =  \mathsf{QN}_{2,u}^o  ,  \\
    \mathsf{QN}_{3,u} &= \sum_{u'\in\mathcal{U}\backslash u} p_{u'} \mathbf{b}_{\mathcal{C}_{u}u}^{\mathsf{H}}
    (
    \mathbf{K}_{\mathcal{C}_u}^{\mathfrak{D}} \odot  \mathbf{b}_{\mathcal{C}_{u}u'}  \mathbf{b}_{\mathcal{C}_{u}u'}^{\mathsf{H}} \
    )
    \mathbf{b}_{\mathcal{C}_{u}u}     \\ 
    &\asymp \sum_{u'\in\mathcal{U}\backslash u} \frac{p_{u'}}{M_u^2}   \sum_{r\in\mathcal{C}_u} 
    \mathrm{tr} \mathbf{\Psi}_{ru} (
    \mathbf{K}_{r}^{\mathfrak{D}} \odot \mathbf{\Psi}_{ru'} 
    )   =  \mathsf{QN}_{3,u}^o  ,  \notag  \\
    \mathsf{QN}_{4,u} &= p_{u} \mathbf{b}_{\mathcal{C}_{u},u}^{\mathsf{H}} 
    (
    \mathbf{K}_{\mathcal{C}_u}^{\mathfrak{D}} \odot  ( \mathbf{b}_{\mathcal{C}_{u},u}  \mathbf{b}_{\mathcal{C}_{u},u}^{\mathsf{H}}  )
    )
    \mathbf{b}_{\mathcal{C}_{u},u}  \notag  \\
    &\asymp \frac{2 p_{u} }{M_u^2}  
    \mathrm{tr} \mathbf{\Psi}_{ru} (
    \mathbf{K}_{r}^{\mathfrak{D}} \odot \mathbf{\Psi}_{ru} 
    )   =  \mathsf{QN}_{4,u}^o  , \label{eq:QN4u}
\end{align}
}where showing~\eqref{eq:QN4u} requires Lemma 2 and the others require Lemma 1. From the continuous mapping and dominated convergence theorems~\cite{wagner_2012}, the DE of user data rate is obtained as $R_u\asymp\log_2(1+\gamma_u^o)=R_u^o$ and the average rate is $\mathbb{E}[R_u]\asymp R_u^o$.

We remark that DEs obtained for the denominator of the SIQNR matches its expectation when MRC is used. This partly aligns with the fact that $\mathbb{E}[\log_2(1+A/B)]\approx\log_2(1+\mathbb{E}[A]/\mathbb{E}[B])$ used in the massive MIMO literature~\cite{zhang_2014}.

\section{Optimal DE-Based Clustering}\label{section:optimal_de_based_clustering}
Although, the simple structure of MRC makes it scalable to dense network settings, it becomes crucial to optimize each user's serving cluster, $\mathcal{C}_u$~\cite{jayawardane_2025}. MRC not only lacks interference mitigation, but also is not aware of the quantization distortion in the fronthaul-limited setting or the impact of imperfect channel estimates. Hence, an optimized clustering scheme compensates for all aspects through better coordination.

Optimizing rates $R_u$ for each channel realization yields good results, but this is impractical since the channels may have changed before optimization finishes. As a practical alternative, we choose to optimize over the interval in which the large-scale fading coefficients (LSFCs) remain constant. More precisely, the DE-based rate $R_u^o$ is used as it serves as a good approximation, even in small dimensions~\cite{wagner_2012}. 

For notational convenience, we assume the case of additive white Gaussian noise (AWGN), $\mathbf{\Sigma}_{\mathbf{n}_r}=\sigma^2\mathbf{I}_M$, and spatially uncorrelated channels, $\mathbf{\Psi}_{ru}=(1-\tau^2)\beta_{rt}\mathbf{I}_M$ and $\tilde{\mathbf{\Psi}}_{ru}=\tau^2\beta_{rt}\mathbf{I}_M$, where $\sigma^2$ is the noise power, $\tau$ is the channel imperfection factor, and $\beta_{ru}$ denotes the LSFC containing pathloss and shadowing. The assumptions are for cleaner presentation and do not fundamentally impact the analysis.

Observe that,  since the channels and noise across APs are uncorrelated, each component of $\gamma_u^o$ can be broken up into contributions of each AP; for example, $\mathbf{\Psi}_{\mathcal{C}_uu}$ is block diagonal. The AP-UE association index is defined as $s_{ru}\in\{0,1\}$ to represent user's serving cluster where $s_{ru}=1$ means $r\in\mathcal{C}_u$ and vice versa. The DE of SIQNR $\gamma_u^o$ without pre-determined clusters can be formulated as
\begin{equation}
    f_u(\mathcal{S}_u) = p_u M (1-\tau^2) \cdot  \frac{(\sum_{r\in\mathcal{R}} a_{ru} s_{ru} )^2}{ \sum_{r\in\mathcal{R}} b_{ru} s_{ru}  }    ,    \label{eqn:siqnr_de}
\end{equation}
\begin{align}
    a_{ru} &= \beta_{ru} , \label{eqn:a_ru} \\
    b_{ru} &= \sigma^2 (1+\kappa_r)\beta_{ru} 
    + p_u (\tau^2 + \kappa_r(2-\tau^2) ) \beta_{ru}^2  \notag  \\
    &\hspace{5.5em}
    + \textstyle\sum_{u'\in\mathcal{U}\backslash u} p_{u'} (1+\kappa_r) \beta_{ru} \beta_{ru'}  , \label{eqn:b_ru}
\end{align}
where $\mathcal{S}_u$ is the set of association variables for user $u$, with $f_u(\mathcal{S}_u)$ defined as zero for an empty serving cluster.

From~\eqref{eqn:siqnr_de}, we note that the user's rate is only a function of the user's cluster, i.e., $f_u$ is only dependent on $\mathcal{S}_u$. Hence, optimizing the network sum data rate is equivalent to optimizing the SIQNR for each user and is formulated as
\begin{subequations}
    \begin{align}
    \mathrm{(P1):} \; \max_{\mathcal{S}_u} \quad &f_{u}(\mathcal{S}_u) \\
    \mathrm{s.t.}\quad\, & s_{ru} \in \{0,1\}, \forall r\in\mathcal{R}  . 
\end{align}
\end{subequations}
We propose to solve the problem by first relaxing the binary feasible set into its convex hull to obtain
\begin{subequations}
    \begin{align}
    \mathrm{(P2):} \; \max_{\mathcal{S}_u} \quad &f_{u}(\mathcal{S}_u) \\
    \mathrm{s.t.}\quad\, & 0\le s_{ru} \le 1 , \forall r\in\mathcal{R}  . 
\end{align}
\end{subequations}
Although (P2) is a non-convex problem, the structure can be exploited to find the global optimum described as follows.

\textit{Theorem 1:} Consider a particular quadratic-over-linear ratio maximization problem with box constraints as
\begin{subequations}
    \begin{align}
    \max_{\{ s_r \}_{\forall r}} \quad & \frac{(\sum_{r} a_{r} s_{r} )^2}{ \sum_{r} b_{r} s_{r}  } \\
    \mathrm{s.t.}\quad\ & 0\le s_{r} \le 1 , \forall r  , 
\end{align}
\end{subequations}
where $a_r,b_r>0,\forall r$, and the objective defined as zero for the trivial solution when $s_r=0,\forall r$.

Define an ordering metric $\rho_r\triangleq b_r/a_r$, a permutation $\pi(\cdot)$ such that $\rho_{\pi(i-1)}\le\rho_{\pi(i)}$, and ordered sums
\begin{equation}
    A_k\triangleq\textstyle\sum_{i=1}^k a_{\pi(i)}, \quad B_k\triangleq\sum_{i=1}^k b_{\pi(i)}.
\end{equation}

Then, there exists a globally optimal solution 
\begin{align}
    s_{\pi(i)}^* &= 
    \begin{cases}
      1, &  i\le k^*,  \\
      0, &  i > k^*,  \\
    \end{cases} \\
    k^* &= \arg\max_{k\in\mathcal{K}} \frac{A_{k}^2}{B_{k}}  , \quad \mathcal{K}=\{1,\cdots,R\} ,
\end{align}
with $R \triangleq|\{s_r\}_{\forall r}|$, the size of the variable set.

\textit{Proof:} The proof is given in Appendix~\ref{appendix:lemma_3}.   \hfill $\blacksquare$

Using Theorem 1, a globally optimal solution can be obtained for (P2). For each user, the ordering metric is $\rho_{ru}=b_{ru}/a_{ru}$, the permutation is $\pi_u(\cdot)$, and the ordered sum is
\begin{equation}
    A_{u,k} = \textstyle\sum_{i=1}^k a_{\pi_u(i)}, \quad B_{u,k}=\sum_{i=1}^k b_{\pi_u (i)} .
\end{equation}
The global optimal solution is obtained as
\begin{align}
    s_{\pi_u(i)u}^* &= 
    \begin{cases}
      1, &  i\le k_u^*,   \label{eqn:optimal_clustering}  \\
      0, &  i > k_u^*,  \\
    \end{cases} \\
    k_u^* &= \arg\max_{k\in\mathcal{K}} \frac{A_{u,k}^2}{B_{u,k}}  , \quad \mathcal{K}=\{1,\cdots,|\mathcal{R}|\} .
\end{align}
Note that the optimal solution \(\mathcal{S}_u^*\) of (P2) lies in the feasible set of (P1). Hence, the solution obtained from~\eqref{eqn:optimal_clustering} is also globally optimal for (P1). The reason is that (P2) maximizes a convex objective over a convex constraint set, so a global optimum is attained at an extreme point of the feasible set. These extreme points are exactly the binary feasible points of (P1). However, it is important to note that the global optimality is achieved with respect to the DE-based rates, but not to instantaneous rates per channel realization or average rates.

There are two aspects to the complexity of the proposed DE-based clustering scheme. First, it only needs to be executed once over each interval during which the LSFCs remain constant, i.e., much less frequently than small-scale channel variations. Second, it attains the global optimum of a combinatorial problem with complexity $\mathcal{O}(|\mathcal{R}|\log|\mathcal{R}|)$ for sorting and $\mathcal{O}(|\mathcal{R}|)$ for searching. 

\section{Numerical Results and Discussions}
In this section, we provide numerical results to compare the average sum data rate and the DE-based approximations as well as to illustrate the effectiveness of the proposed clustering scheme. Some important parameters chosen are: The APs and UEs are uniformly distributed in a square region of 400m per side, with wraparound. The AWGN and spatially uncorrelated channel model is used as in Section~\ref{section:optimal_de_based_clustering}. A Gaussian quantizer with $\Gamma_q=1$ and channel imperfection factor of $\tau=0.2$ is considered. All users transmit at the same power with $p_u=23$ dBm. Interference is significantly stronger than noise with pathloss being $\beta_{ru}=(1+d_{ru}/0.392)^{-3.76}$, where $d_{ru}$ is the distance, and noise power is $\sigma^2=-106$ dBm.

First, we show that the DE-based approximations are accurate by comparing it with the empirical average sum rate. For the average rate, Monte-Carlo (MC) simulation is used and averaged over $1000$ different channel realizations and topologies. For the DEs, only topological averages are made as they do not depend on small-scale fading.

\begin{figure}
	\centering
	\includegraphics[width=1.0\linewidth]{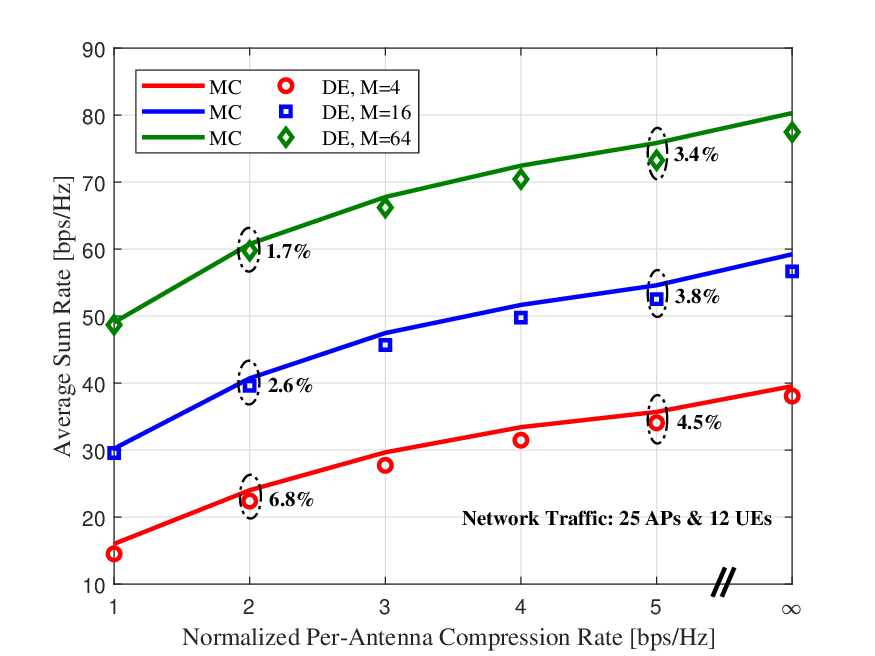}
    \captionsetup{font=small}
	\caption{Sum of user data rates, obtained using MC and DE, versus pre-antenna compression rate, $L_r/M$, with each user served by $|\mathcal{C}_u|=5$ nearest APs. }
	\label{figure:plot1_vary_fronthaul}
\end{figure}

From Fig.~\ref{figure:plot1_vary_fronthaul}, the relative gap between the average sum rates and corresponding DEs decrease as number of antennas at each AP, $M$, increases; this suggests equivalence in the asymptotics. Even with small $M$, the DE is quite accurate. More importantly, Fig.~\ref{figure:plot1_vary_fronthaul} justifies use of DEs as the optimization objective as it accurately captures the trend of the sum rates.

Next, we illustrate the effectiveness of the proposed DE-optimal clustering ($\gamma^o$-Optimal) by comparing to the following benchmarks:
\begin{itemize}
    \item Greedy LSFC-based selection with fixed cluster size ($\beta$-Fixed): Select the $|\mathcal{C}_u|$ APs with largest LSFCs. $|\mathcal{C}_u|$ is predetermined and the same for all $u$ and denoted by ``-Fixed" in the notation.
    \item Greedy local SIQNR-DE-based selection with fixed cluster size (Local$\gamma^o$-Fixed): Select the APs with top-$|\mathcal{C}_u|$ largest DE of local SIQNR, which is a psuedo-quantity as if user $u$ is only processed at AP $r$ and obtained as $\gamma_{u|r}^o=f_u(s_{ru}=1,s_{r'u}=0,\forall r'\in\mathcal{R}\backslash r)$ using~\eqref{eqn:siqnr_de}.
    \item Greedy local SIQNR-based selection with fixed cluster size (Local$\gamma$-Fixed): Same as Local$\gamma^o$-Fixed but for every realization. The instantaneous local SIQNR is obtained as $\gamma_{u|r}=\gamma_r|_{\mathcal{C}_u=\{r\}}$ using~\eqref{eqn:siqnr}.
    \item Optimal clustering that maximizes \textit{instantaneous} SIQNR for each channel realization ($\gamma$-Optimal): Maximizing~\eqref{eqn:siqnr} for every realization using exhaustive search.
\end{itemize}
We emphasize that the above benchmarks are strong baselines, as they improve upon existing literature, e.g.,~\cite{jayawardane_2025}, by accounting for quantization distortion and channel imperfections. The final benchmark represents the best clustering performance attainable without complexity requirements.

\begin{figure}
	\centering
	\includegraphics[width=1.0\linewidth]{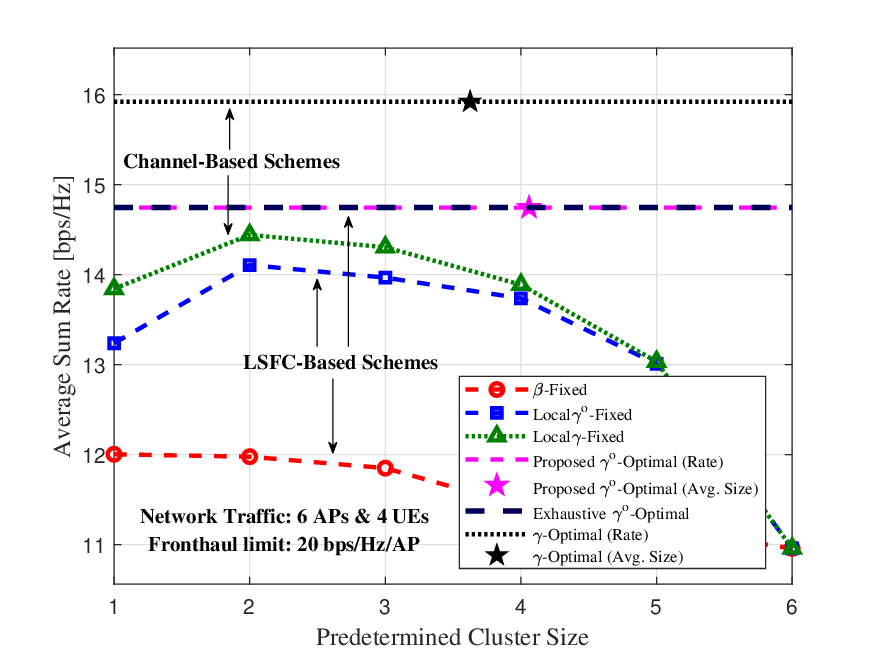}
    \captionsetup{font=small}
	\caption{Average sum rate, obtained from MC, versus cluster schemes. }
	\label{figure:vary_schemes}
\end{figure}

\begin{figure}
	\centering
	\includegraphics[width=1.0\linewidth]{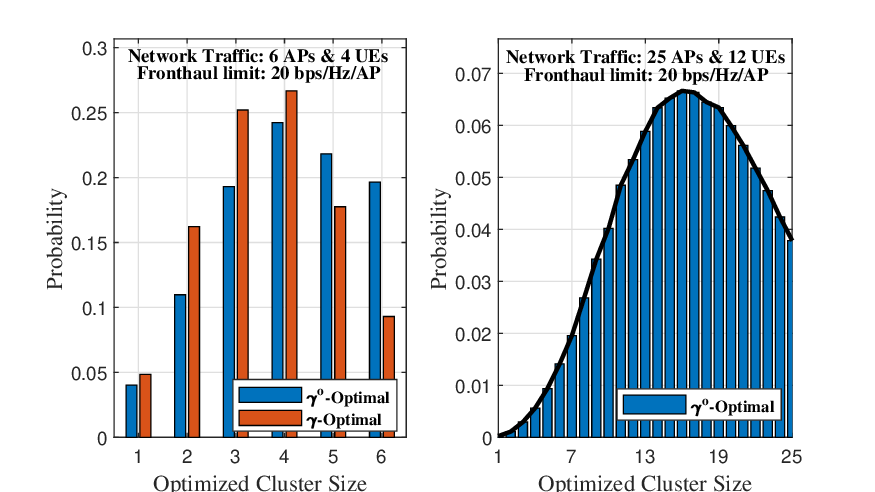}
    \captionsetup{font=small}
	\caption{Probability mass function for optimized cluster sizes, $|\mathcal{C}_u^*|$. }
	\label{figure:pmf}
\end{figure}






In Fig.~\ref{figure:vary_schemes}, a relatively small network is considered such that it is possible to execute the exhaustive search. Each AP has $M=4$ antennas. The figure compares the performance of our proposed scheme to two LSFC-based benchmarks, which are updated at the same frequency as the proposed method. Due to interference, $\beta$-Fixed, a common clustering heuristic when considering MMSE combiners, underperforms when MRC is used. Another heuristic, Local$\gamma^o$-Fixed, offers better performance but it requires searching for best cluster size and sorting the local SIQNR, which is superseded by the proposed method as the latter has similar complexity but provides the global optimum under the DE-based objective.

Two channel-based benchmarks, that are updated every channel realization which is significantly more frequent, are also compared. The globally optimal clustering approach, $\gamma$-Optimal, is obtained through exhaustive search but is impractical for large networks as it scales exponentially for every channel instance. On the other hand, our proposed scheme is polynomial-time and updated less frequently but only penalized by 7.4\% in terms of the average sum rates. 

Another polynomial-time heuristic, Local$\gamma$-Fixed, underperforms despite the fact that it is updated for every small-scale fading. This suggests the importance of dynamic service clusters for each user when using MRC as it implicitly manages the interference without complicating the combiner structure. The advantage is further illustrated in Fig.~\ref{figure:pmf} which shows that the optimized cluster sizes, $|\mathcal{C}_u^*|$, vary significantly across users. This indicates that spatial diversity is utilized adaptively rather than wasted as under fixed-size clustering.


\section{Conclusions}\label{section:conclusion}
This paper investigates scalable distortion-aware clustering for uplink fronthaul-limited cell-free networks with imperfect channel knowledge and MRC receivers. By leveraging large system analysis, a deterministic equivalent expression of the average network sum rate, under fronthaul distortions, depending only on the LSFCs is derived. This makes the clustering optimization more practical, since LSFC-based updates are required far less frequently than updates based on small-scale fading. A polynomial time scheme is then developed to solve the resulting DE-based problem, and it attains global clustering optimum when restricted to LSFCs. Although low-complexity MRC suffers from interference and fronthaul distortion, the results show that optimized clustering yields meaningful performance gains even when only LSFCs are available.


\appendices

\section{Proof of Lemma 2}\label{appendix:lemma_2}
Let $a_i=[\mathbf{A}_N^{\mathfrak{D}}]_{ii}$, $d_i=[\mathbf{B}_N]_{ii}$, and $b_i=\sqrt{N}[\mathbf{B}_N^{1/2}\mathbf{x}_N]_{i1}$ such that $b_i\sim\mathcal{CN}(0,d_i)$. Set $\xi_i=a_i(|b_i|^4-2d_i^2)$, the gap between the stochastic expression and the proposed DE is
\begin{equation}
    \Delta_N=\frac{1}{N^2}\sum_{i=1}^N \xi_i.
\end{equation}
The gap is zero mean since $\mathbb{E}[|b_i|^4]=2d_i^2$ and it can be shown that $\mathbb{E}[\xi_i^2]=20a_i^2d_i^4$ from $\mathbb{E}[|b_i|^8]=24d_i^4$. The variance of the gap is bounded by 
\begin{equation}
    \mathbb E[\Delta_N^2]
    \le \frac{1}{N^3}\sum_{i=1}^{N} 20a_i^2d_i^4  ,
\end{equation}
using Cauchy-Schwarz inequality and without requiring independence between $b_i$.

Since $\mathbf{A}_N^{\mathfrak{D}}$ and $\mathbf{B}_N$ are uniformly bounded, there exists a finite constant $C$ such that $a_i^2d_i^4 \le C, \forall i$ and
\begin{equation}
    \mathbb{E}[\Delta_N^2] \le \frac{20C}{N^2} .
\end{equation}

Choosing $\epsilon_N=N^{-k}$ and from Chebyshev's inequality, 
\begin{equation}
    \Pr \left(
    |\Delta_N| \ge \epsilon_N
    \right) \le 
    \frac{\mathbb{E} \left[ \Delta_N^2 \right]}{\epsilon_N^2} \le
    20C N^{2k-2} .
\end{equation}
The sum of probability of the sequence of events $\{\{ |\Delta_N| \ge \epsilon_N \}\}_{N=1}^{\infty}$ is a converging series by choosing $0<k<1/2$. The sum is
\begin{equation}
    \sum_{N=1}^{\infty} \mathrm{Pr}\left( |\Delta_N| \ge \epsilon_N  \right) \le  
    \sum_{N=1}^{\infty} 20C N^{2k-2} < \infty ,
\end{equation}
is finite. By the Borel-Cantelli lemma~\cite{wagner_2012}, 
\begin{equation}
    \mathrm{Pr}\left( \limsup_{N\rightarrow\infty} \left\{ |\Delta_N| \ge \epsilon_N \right\} \right) = 0   ,
\end{equation}
which suggests $|\Delta_N| < \epsilon_N$ almost surely for sufficiently large $N$. As $N\rightarrow\infty$, the gap goes to zero almost surely which completes the proof.


\section{Proof of Theorem 1}\label{appendix:lemma_3}
Define the variable transform $x_r\triangleq a_r s_r$ such that $0\le x_r\le a_r$, and its sum of all $r$ as $U$. The problem is equivalent to
\begin{subequations}
    \begin{align}
    \max_{\{ x_r \}_{\forall r},U} \quad & \frac{U^2}{ \sum_{r} \rho_{r} x_{r}  }  \label{eqn:appendix_reform_obj1} \\
    \mathrm{s.t.}\quad\ & 0\le x_{r} \le a_r , \forall r  ,    \\
    & U= \textstyle\sum_{r} x_r   .
\end{align}
\end{subequations}
We remark that, for fixed $U$, the objective~\eqref{eqn:appendix_reform_obj1} can be maximized greedily by filling up $x_{\pi(i)}$ in ascending order of $\rho_{\pi(i)}$ until the sum reaches $U$ because the denominator is minimized. In other words, there exists an index $k$ such that
\begin{equation}\label{eqn:appendix_optimal_x}
    x_{\pi(i)}^* = 
    \begin{cases}
      a_{\pi(i)}, &  i < k ,  \\
      U - A_{k-1} , & i=k,   \\
      0, &  i > k .  \\
    \end{cases}
\end{equation}
The denominator can be rewritten in terms of $U$ and $k$ as
\begin{align}
    \textstyle\sum_{r} \rho_{r} x_{r} &= \textstyle\sum_{i=1}^k \rho_{\pi(i)} x_{\pi(i)} \notag \\
    &= (B_{k-1} - \rho_{\pi(k)} A_{k-1} ) + \rho_{\pi(k)} U   .
\end{align}
In addition, $U$ and $k$ are related as $k$ decides which window $U$ is located in, i.e., $U\in[A_{k-1},A_{k}]$. As a result, the problem is equivalent to 
\begin{subequations}
\begin{align}
    \max_{U,k}\quad &
    \frac{U^2}{(B_{k-1}-\rho_{\pi(k)}A_{k-1})+\rho_{\pi(k)}U}
    \label{eqn:appendix_reform_obj2} \\
    \mathrm{s.t.}\quad &
    k\in\{1,\ldots,R\}, \qquad U\in[A_{k-1},A_k].
\end{align}
\end{subequations}  
where the empty sum $A_0,B_0\triangleq0$. Observe that, for fixed $k$, the objective~\eqref{eqn:appendix_reform_obj2} is convex so the maximizing points are at extreme points, i.e., $U^*=A_{k-1}$ or $U^*=A_{k}$. Hence, the global optimal solution can be expressed as
\begin{equation}
    k^* = \arg\max_{k\in\mathcal{K}} \frac{A_{k}^2}{B_{k}}  , \quad \mathcal{K}=\{1,\cdots,R\}  ,
\end{equation}
and corresponds to $U^*=A_{k^*}^2$. Plugging in~\eqref{eqn:appendix_optimal_x} and mapping back to $s_r$, the global optimal solution to the original problem is
\begin{align}
    s_{\pi(i)}^* &= 
    \begin{cases}
      1, &  i\le k^*,     \\
      0, &  i > k^*,  \\
    \end{cases} 
\end{align}
which completes the proof.

\bibliography{biblio}
\bibliographystyle{ieeetr}

\end{document}